\documentclass[oneside,12pt]{article}

    \title{Time and Dirac Observables in Friedmann cosmologies }

    \author{ Hossein Farajollahi \footnote{Department of Physics,
    Faculty of Science, The University of Guilan,
    Email: hosseinf@guilan.ac.ir  }}

     \date{}

\begin{document}
    \maketitle

    \begin{abstract}

A cosmological time variable is emerged from the Hamiltonian
formulation of Friedmann model to measure the evolution of
dynamical observables in the theory. A set of observables has been
identified for the theory on the null hypersurfaces that its
evolution is with respect to the volume clock introduced by the
cosmological time variable.

$\bf{Keywords}$: Cosmology, Relativity, FRW, Friedmann, Time, Dirac,
Hamiltonian, Observable

\end{abstract}

    \newpage

    \def\be{\begin{equation}}
    \def\ee{\end{equation}}
    \def\bea{\begin{eqnarray}}
    \def\eea{\end{eqnarray}}
    \def\M{{{\cal M}}}
    \def\bdy{{\partial\cal M}}
    \def\w{\widehat}
    \def\n{\widetilde}
    \def\real{{\bf R}}

    \section{Introduction}

Research in quantum gravity may be regarded as an attempt to
construct a theoretical scheme in which ideas from General
Relativity and quantum theory are reconciled. However, after many
decades of intense work we are still far from having a complete
quantum theory of gravity. Any theoretical scheme of gravity must
address a variety of conceptual issues including the problem of time
and identification of dynamical observables. There are many program
that attempt to address the above mentioned problems including
canonical quantum gravity.

It is well know that some of the issues such as time and observables
in quantum gravity have their roots in classical general relativity;
in such cases it seems more reasonable to identify and perhaps
address the problem first in this context.The classical theory of
gravity is invariant under the group of  Diff (${\cal M}$) of
diffeomorphisms of the space-time manifold ${\cal M}$ . To be more
specific, the theory is invariant under time reparametrization and
spacial diffeomorphism. This goes against the simple Newtonian
picture of the a fixed and absolute time parameter. The classical
theory, while itself free from problems relating to the definition
and interpretation of time, contains indications of problems in the
quantum theory, where the absence of a time parameter is hard to
reconcile with our everyday experience. In fact, one can see that in
the Hamiltonian formulation of classical general relativity, time is
suppressed from the theory. There are many proposals for dealing
with this question which generally involve a re-interpretation of
the usual notion of time ( see \cite{Isham} for an overview of these
proposals).

 Identification of dynamical observable for the theory is another fundamental issue
that has its roots in classical formulation of general relativity
and directly related to the issue of time. The problem of evolving
of a dynamical system from initial data is known as the Cauchy
problem or initial value problem \cite{Inverno} and in General
Relativity is naturally addressed using the 3+1 ADM representation.
In the Arnowitt-Deser--Misner (ADM) approach, the spatial
hypersurface $\Sigma(t)$ is assumed to be equipped with a space-like
3-metric $\gamma_{ij}$ ($i, j$ runs from $1$ to $3$) induced from
space-time metric $g_{\mu \nu}$($\mu ,\ \nu $ runs from $0$ to $3$).
Einstein's equations are of course covariant and do not single out a
preferred time with which to parametrise the evolution.
Nevertheless, we can specify initial data on a chosen spatial
hypersurface $\Sigma$ , and if $\Sigma$ is Cauchy, we can evolve
uniquely from it to a hypersurface in the future or past. The issue
of specification of initial or final data on Cauchy hypersurfaces
has been discussed in many papers; for example, see \cite{Hawking}.

 An alternative approach to Cauchy problem is known as characteristic
 initial value problem in which one may fix the initial data on null
 hypersurfaces rather than
spatial hypersurfaces. There are reasons to motivate us using null
boundaries in formulating general relativity. For a summary one
may look at the \cite{Komar}\cite {Dautcourt} \cite {hughhossein}
\cite {Penrose} \cite {Sachs} \cite {Ellis} \cite {Bondi}.

 In addition, the approach of setting the final data on a null hypersurface is
 essential if we are interested in a theory such as quantum theory
 that observations made by a single localized observer who can collect
 observational data only from that subset of space-time which lies in the
 causal past\cite{hughhossein}.

Studying cosmological models instead of General relativity helps us
to overcome the problems related to the infinite number of degrees
of freedom in the theory and pay more attention to the issues
arising from the time reparametrization invariance of the theory;
such as the identification of a dynamical time and also construction
of observables for the theory\cite{palii}.

There are many general homogeneous (but anisotropic) cosmological
models such as the Kantowski-Sachs models and the Bianchi models.
However, in this paper we consider Friedmann-Robertson-Walker (FRW)
cosmologies for simplicity. The standard FRW universe are of course
one special example. In that case we assume that our universe filled
with scalar massless matter field which simply has two
minisuperspace coordinates, $\{a,\phi\}$, the cosmic scale factor
and the scalar field. The conventional Hamiltonian formulation of
this model based on Dirac and ADM procedure of general relativity is
developed.

The main feature of the Hamiltonian theory of gravity is the
presence of nonphysical variables and constraints due to the
diffeomorphism invariance of the theory. As mentioned, This is in
turn an obstacle to the problem of identification of a time
parameter to measure physical quantities such as cosmological
observables (the Hubble law and red shift) and the Dirac observables
in the Hamiltonian description of the classical and quantum
cosmologies. One of the possible solutions of these problems in the
Hamiltonian approach discussed in this paper is to reduce the
original theory reparametrization-invariantly by the explicit
resolving of the first class constraints to get an equivalent
unconstrained system. In this approach one of the variables of the
extended phase space converts into the dynamic evolution parameter
that plays the role of a cosmological time \cite{hughhossein} in the
theory. Thus, instead of the extended phase space and the initial
action invariant under reparametrizations of the coordinate time, we
obtain the reduced phase space which contains only the matter field
described by the reduced Hamiltonian.

In this paper in section two, the Hamiltonian formulation of a
simple reparametrization invariant model has been presented. A
reduced Hamiltonian and a time variable has been emerged from this
model. In section three, we apply the Hamiltonian reduction
developed in section two to FRW model when a massless scalar field
minimally coupled to gravity .

In section four, a discussion of Dirac observables in general
relativity is given. The 'Rovelli's constants of motion'
\cite{Rovelli} have also been discussed. In section five Dirac
observables on the null hypersurface of a single localized observer
for FRW cosmologies is identified. These observables are similar to
Rovelli's constants of motion on the null hypersurfaces
\cite{hughhossein}. The evolution of these observables is with
respect to time variable obtained from massless scalar field coupled
to the gravity.

\section{A simple parametrized model}

To construct a reduced phase space with a reduced Hamiltonian for a
time reparametrization invariant system let us begin with a simple
toy model in classical mechanics. The case of a one dimensional
motion of a particle with the action given by \be
S[x,\sigma]=\frac{1}{2}\int_{t_0}^{t_f}
(N^{-2}\dot{x}^2-v(x)-N^{-2}\dot{\sigma}^2)Ndt,
\label{eq:particleaction}
 \ee
where $N^2$ is contravariant metric and $x(t)$, $\sigma(t)$ and
$N(t)$ are the independent configuration variables for the particle.
With the $d\tau=Ndt$ to be the proper time interval and $v(x)$ the
potential, one can rewrite the action as \be
S[x,\sigma]=\frac{1}{2}\int_{\tau(t_0)}^{\tau(t_f)}
[(\frac{dx}{d\tau})^2-v(x)-(\frac{d\sigma}{d\tau})^2]d\tau.
 \ee
With the gauge fixing, $N=1$, the dynamics is unique which is out of
our interest. Without gauge fixing, i.e. allowing $N(t)$ to vary,
according to the Dirac prescription the generalized Hamiltonian
dynamics for the action (\ref{eq:particleaction}) takes place on the
phase space spanned by the three canonical pairs $(x, p_x)$,
$(\sigma, p_\sigma)$ and $(N, p_N)$.

Since $\sigma$ is a dynamical variable (while $\tau$ not) and has a
simple dynamics i.e. $\sigma=\alpha\tau+\beta$ (on shell), one can
use $\sigma$ (rather than $\tau$) to parametrize $x$ and also it may
be considered as a clock time to make measurement.

The Euler-Lagrange equation for the dynamical variable $x$ with
respect to $\tau$ and $\sigma$ are:

$$\frac{d^2x}{d\tau^2}=-\frac{\partial v}{\partial x}, \ \ \ \
\frac{d^2x}{d\sigma^2}=-\frac{1}{\alpha^2}\frac{\partial
v}{\partial x},$$ where $\alpha$ is a measurable constant. Thus,
one may consider evolution of $x(\sigma)$ with respect to
(measurable) clock time $\sigma$ instead of $\tau$.

The momenta associated with the dynamical variables are
$$p_x=N^{-1}\dot{x},\ \ \  p_\sigma=-N^{-1}\dot{\sigma}, \ \ \  \
\ \ p_N=0.$$ Since the action does not explicitly depends on the
variable $\dot{N}$, the vanishing momenta $p_N$ is primary
constraint, $$p_N\approx 0.$$

The canonical Hamiltonian then is \be H_0=p_x\dot{x}+p_\sigma
\dot{\sigma}-L=\frac{1}{2}N[p_x^2-p_\sigma^2+v(x)] \ee
 with the
total Hamiltonian

$$H_T=\lambda_Np_N-H_0=\lambda_Np_N+\frac{1}{2}N[p_x^2-p_\sigma^2+v(x)].$$
 Following Dirac procedure, to ensure that the primary constraint
is preserved with time evolution, we also require that
$\dot{p}_N\approx 0.$ which gives us the secondary constraint
$$0\approx \frac{1}{2}[p_x^2-p_\sigma^2+v(x)]=\cal{H},$$ and
therefore the total Hamiltonian now reads \be H_T=\lambda_N
p_N+N\cal{H}, \ee where the variables $\lambda_N$ and $N$ in
Hamiltonian are Lagrange multipliers. Now, the equations of motion
for our system are

\bea \dot{x}=Np_x  ,\ \ \ \ \    \dot{p}_x=\frac{1}{2}Nv',\\
\dot{\sigma}=-Np_\sigma   ,\ \ \ \ \     \dot{p}_\sigma=0,\\
\dot{N}=\lambda_N, \ \ \ \ \    \dot{p}_N=\lambda_N,  \eea
which
accompanied with the two first class constranits (FCC):

\be
{\cal H}\approx 0, \ \ \ \ \ \  p_N\approx 0. \ee

It is easy to check that according to Dirac procedure $x(t)$ which
is the value of $x$ for a given value of $t$ is not eligible to be
an observable since $\{H, x(t)\}\neq 0.$ it means that specifying t
does not identify a special point on the trajectory as the
parametrization of the trajectory is not fixed. However, $x(\sigma)$
which is the value of $x$ for a given value of $\sigma$ is an
eligible observable since $\{H, x(\sigma)\}=0$. It gives us particle
location when clock says for example $3:20$. Once measured and
recorded, stays fixed for all time. (historical record!)

Among the dynamical variables, only $p_\sigma$ is a first class
variables since its poisson brackets with the constraints vanishes.
Thus, one can introduce a new canonical variables for ($\sigma
,p_\sigma $) as \be T_\sigma=\frac{\sigma}{p_\sigma},\ \ \ \ \ \
p_T=\frac{p_\sigma^2}{2},\ee in order to obtain a reduced
Hamiltonian describing the evolution of the particle with respect to
the new dynamical time variable $T_\sigma$.

In terms of the new variables the total Hamiltonian is \be
H_T=\lambda_N p_N+Np_T, \ee and one can divide the equations of
motion into two parts:1) for the canonical pairs $(N, p_N)$,
$(T_\sigma, p_T)$ with a dependency on the Lagrange multiplier
$\lambda(\tau)$, \bea  \dot{T}_\sigma=N ,\ \ \ \ \ \dot{p}_T=0,\\
\dot{N}=\lambda_N, \ \ \ \ \ \dot{p}_N=-p_T, \eea which
constrained by $p_T=0$. 2)for the canonical pair $(x, p_x)$ \be
\dot{x}=0 ,\ \ \ \ \ \dot{p}_x=0 \ee which have a unique solution
with no constraint. The reduced Hamiltonian that governs the
particle evolution in time $T_\sigma$ then is

$$H(x)=\frac{1}{2}[p_x^2+v(x)].$$

Note that although the dynamical time $T_\sigma$ does not commute
with the constraints and so is not a first class variable but its
momenta $p_T$ is a first class variable and so eligible to be
considered as a time variable to measure the passage of time.

Alternatively one can reduce the theory in terms of the coordinate
$x$ by performing the canonical transformation on $x$

\be T_x=\int dx (2\Pi_x+v)^{-1/2},\ee \be
\Pi_x==\frac{1}{2}[p_x^2-v]\ee and thus the reduced Hamiltonian that
describes the evolution of the variable $\sigma$ in time $T_x$ is

\be H(p_\sigma)=\frac{p_\sigma^2}{2}.\ee

Once again only those new canonical variables are eligible to be
considered as dynamical time that their associated momenta are first
class variables.

\section{FRW model with scalar field minimal coupling to gravity}

We begin with the line element for the FRW model in spherical
coordinates
\be ds^2=-N^2(t)dt^2+a^2(t)h_{ij}dx^idx^j,
\label{eq:FRWmetric}\ee where $N(t)$ is the lapse function, $a(t)$
is the cosmic scale factor determines the radius of the universe,
and $h_{ij}$ is the time independent metric of the three-dimensional
maximally symmetric spatial sections \be
h_{ij}dx^idx^j=\frac{dr^2}{1-kr^2}+r^2(d\theta^2+sin^2\theta
d\phi^2) \ee
 of constant curvature ${}^{(3)}R(h_{ij})=-6k$,
$k=0,\pm 1.$

 Inserting the metric  (\ref{eq:FRWmetric}) into the action for vacuum
 FRW model in the natural units gives
$c =h = 1$ \be S[g_{\mu\nu}]=\int \sqrt{-g}\ {}^{(4)} R d^4x=\int
dt \int_{\Sigma(t)} d^3x \sqrt{-g}\ {}^{(4)}R.
\label{eq:FRWmetric1}\ee By assuming the spatial homogeneity of
the FRW metric the action (\ref{eq:FRWmetric1}) can be written as

\be S[g_{\mu\nu}]=\int dt \int  3(\frac{a\dot{a}^2}{N}-kNa)d^3x=
V_{(3)}\int 3(\frac{a\dot{a}^2}{N}-kNa)dt,\ee where $V_{(3)}$ is the
volume of the three-dimensional space of constant curvature. The
momenta conjugate with the dynamical variables are

$$\Pi_a=6N^{-1}a\dot{a},$$

$$\Pi_N=0,$$  which $\Pi_N\approx 0$ is a primary constraint.

The canonical Hamiltonian is $$H_0=(\frac{N\Pi_a^2}{12}+kNa),$$
and thus the total Hamiltonian is \be
H_T=\lambda_N\Pi_N+N(\frac{\Pi_a^2}{12}+ka). \ee The secondary
constraint is \be 0\approx {\cal H}=\frac{\Pi^2_a}{12}+ka \ee and
therefore \be H_T=\lambda_N\Pi_N+N\cal{H}.\ee One can check that
our constraints are both FCC. The number of dynamical variables
are four ($a, \Pi_a ; N, \Pi_N)$. Since there are two FCC
constranits, it means that the vacuum FRW model
 has no physical degrees of freedom on the classical level and
 only unphysical degrees of freedom propagate. So in order to have
 some non-trivial observables it is necessary to introduce the
 source matter fields.

The Einstein-Hilbert action for the FRW model for the gravity
minimally coupled to a massless scalar field is given by \be
S[g_{\mu\nu},\phi]=\int_{\Sigma(t)}\int \sqrt{-g}( -\frac{1}{2}\
 {}^{(4)}R+\frac{1}{2}g^{\mu\nu}\partial_\mu \phi
\partial_\nu \phi ) dtd^3x \label{eq:actionintegral},\ee
which by assuming the spatial homogeneity of the scalar field in
FRW metric yields \be S[g_{\mu\nu},\phi]=V_{(3)}\int
[3(\frac{a\dot{a}^2}{N}-kNa)-\frac{a^3}{2N}\dot{\phi}^2]dt
\label{eq:actionintegral1}.\ee

Given the action-integral (\ref{eq:actionintegral}) or
(\ref{eq:actionintegral1}) it is easy to find the canonically
conjugate momentum to the dynamical variable,   \bea
p_a=6N^{-1}a\dot{a},\\ p_\phi=-N^{-1}a^3\dot{\phi}, \eea and \be
p_N=0\ee which is a primary constraint.

The Hamiltonian is

$$H_0=N[\frac{p_a^2}{12a}+3ka-\frac {p_\phi^2}{2a^3}]$$ and thus
the total Hamiltonian is \be
H_T=\lambda_Np_N+\frac{N}{a^3}[\frac{a^2p_a^2}{12}+3ka^4-\frac{p_\phi^2}{2}].
\ee

By the non-degenerate character of the metric ($a\neq 0$), the
secondary constraint can be redefined (choosing $N_1=N/a^3$) \be
0\approx {\cal H}_1=\frac{a^2p_a^2}{12}+3ka^4-p_\phi^2/2 \ee which
shows the separability of the gravitational and the matter source
part in the constraint. Thus, the total Hamiltonian becomes \be
H_T=\lambda_Np_N+N_1{\cal H}_1.\label{eq:totalHamiltonian}\ee

One can check that our constraints are both FCC. The number of
dynamical variables are six,( $a, p_a ;\phi , p_\phi ;  N, p_N)$.
There are two FCC constranits. Thus, there are only two physical
degrees of freedom on the classical level in the FRW model and
 only these two physical degrees of freedom propagate. According to
 the the procedure described in the last section since a unique
 solution cannot be find for the equations of motion one has to
 implement the Hamiltonian reduction to separate the equations of
 motion into the physical and the unphysical ones. For this, let us
 introduce the new canonical variable in order to obtain the
reduced Hamiltonian describing the evolution of the cosmic scalar
factor $a$, \be T_\phi=\int_{\Sigma}
\frac{\phi}{p_\phi}\sqrt{{}^{(3)}h}d^3x,\label{eq:cosmologicaltime}
\ee \be
 \Pi_\phi=\frac{p_\phi^2}{2}.\ee
In terms of the new variables the total Hamiltonian is: \be
H_T=\lambda_N p_N+N_1\Pi_\phi. \ee Similarly, in here, one can
separate the equation of motion into two parts: one for the
canonical pairs $(N, p_N)$, $(T_\phi, \Pi_T)$ with a dependence on
the Lagrange multiplier $\lambda(\tau)$ \bea \dot{T}_\phi=NV_{(3)}
,\ \ \ \ \ \dot{\Pi}_\phi=0\\ \dot{N}=\lambda_N, \ \ \ \ \
\dot{p}_N=-\Pi_\phi \eea constrained by $\Pi_\phi=0$ and second
for the dynamical variables $(a, \Pi_a)$, \be \dot{a}=0 ,\ \ \ \ \
\dot{\Pi}_a=0 \ee which have a unique solution with initial values
free from any constraints. The reduced Hamiltonian that governs
the scale factor evolution in time $T_\phi$ is

$$H(a)=\frac{a^2p_a^2}{12}+3ka^4.$$

 The equation of motion for $T_\phi$ derived from
(\ref{eq:cosmologicaltime}), \be
\frac{dT_\phi}{dt}=\int_{\Sigma(t)} \sqrt{-{}^{(4)}g}d^3x.\ee
implies that $T_\phi(t)$ is just the 4-volume preceding $\Sigma_t$
plus some constant of integration. Integration with respect to
$t$, this means that, the change of the time variable equals the
four-volume enclosed between the initial and final hypersurfaces,
which is necessarily positive. This time variable, $T_\phi(t)$ may
be regarded as a cosmological time variable, as it continuously
increasing along any future directed time-like curve
\cite{Hossein}. Assuming that the scalar field is spatially
homogeneous and monotically increasing along the world line
observable normal to the spatial hypersurfaces, one therefore my
consider $T_\phi(t)$ as a monotonically increasing function along
any classical trajectory and so can indeed be used to parametrise
this trajectory \cite{hughhossein}.
 This describes the
evolution of geometry with respect to the dynamical time
constructed from scalar field.

Alternatively, by reformulating the theory in terms of scalar
field one may construct a dynamical time from geometry. To find
the dynamics of the scalar field we perform the canonical
transformation on the scalar factor \be T_a=\int_{\Sigma(t)}
\sqrt{{}^{(3)}h}d^3x\int
\frac{a^2da}{(\frac{\Pi_a}{3}-a^4k)^{1/2}},
\label{eq:cosmologicaltime1}\ee \be
\Pi_a=a^2[\frac{p_a^2}{12}+3ka^2].\ee

One can also show that the dynamical time constructed from sclar
factor $a$ is a $4$-volume preceding $\Sigma(t)$.

\section{Dirac observables in General Relativity}

General Relativity, like many other field theories, is invariant
with respect to a group of local symmetry transformations
\cite{Marolf}. The local symmetry group in General Relativity is
the group Diff (${\cal M}$) of diffeomorphisms of the space-time
manifold ${\cal M}$.

In General Relativity, Dirac observables \cite{Dirac} must be
invariant under the group of local symmetry transformations. The
Hamiltonian constraint and momentum constraint in General
Relativity are generators of the symmetry transformations, and so
a function $\Psi$ on the phase space is a Dirac observable, ${\it
iff}$
 \be
\{\Psi,{\cal H}\}=\{\Psi,{\cal H}_i \}= 0,
 \ee
at all points $x \in {\cal M}$, where ${\cal H}$ and ${\cal H}_i $
are Hamilltonian and momentum constraints in general relativity.

Such observables are necessarily constants of motion. They are
invariant under local Lorentz rotations ${\it SO(3)}$ and ${\it
Diff}\Sigma$   (as well as ${\it SO (1, 3)}$).

The above criteria for observables in relativity appear to rule out
the existence of local observables if locations are specified in
terms of a particular coordinate system. Indeed, it might appear
that one would be left with only observables of the form \be
\Psi=\int \psi(x)\sqrt{-g} \ d^4 x,
 \ee
where  $\psi(x)$ is an invariant scalar as for example $R$, $R^2$ ,
$R^{\mu\nu} R_{\mu\nu}$, etc . While such observables clearly have
vanishing Poisson brackets with all the constraints, they can not be
evaluated without full knowledge of the future and past of the
universe. While this may be deducible in principle from physical
measurements made at a specific time, it is well beyond the scope of
any real experimenter.

However, in reality, observations are made locally. We therefore
ought to be able to find a satisfactory way to accommodate local
observables within General Relativity. In particular, we would
like to be able to talk about observables measured at a particular
time, so that we can discuss their evolution. Local observables in
classical or quantum gravity must be invariant under coordinate
transformations. The difficulty in defining local observables in
classical gravity is that diffeomorphism invariance makes it
difficult to identify individual points of the space-time manifold
\cite{hughhossein}\cite{Camelia}.

It is fairly easy to construct observables which commute with the
momentum constraints. Such observables can be expressed as
functions of dynamical variables on the spatial hypersurfaces.
However, according to the Dirac prescription, observables must
also commute with Hamiltonian constraint.

In a slightly different formalism, Rovelli addressed the problem
by introducing a Material Reference System (${\it MRS}$)
\cite{Rovelli}. By ${\it MRS}$, Rovelli means an ensemble of
physical bodies, dynamically coupled to General Relativity that
can be used to identify the space-time points.

Rovelli's observables can be interpreted as the values of a
quantity at the point where the particle is and at the moment in
which the clock displays the value $t$. However $t$ itself is not
an observable, even though its conjugate momentum is constant
along each classical trajectory.

Rovelli's observables are constant of motion since they commute
with Hamiltonian and momentum constraints, while evolving with
respect to the clock time $t$.

Rovelli's observables are functions defined on spatial
hypersurfaces. He assumes the space-time has a topology $\Sigma
\times R$ where $\Sigma$ is a compact spatial hypersurface and R
is the real time. In order to have evolution into the future or
past the spatial hypersurface must be a Cauchy hypersurface. This
makes sense if the underlying space-time is assumed to be globally
hyperbolic.

As discussed, one may fix the initial data on null hypersurfaces
rather than spatial hypersurfaces. In General Relativity it is
natural to work with a foliation of space-time by space-like
hypersurfaces, as this reflects the older Newtonian idea of a
3-dimensional universe developing with time. This seems close to
our experiences and is easy to visualize. In particular The
approach of setting the final data on a null hypersurface is
essential if we are interested in a theory such as quantum theory
that observations made by a single localized observer who can
collect observational data only from that subset of space-time
which lies in the causal past.

\section{ Dirac observables in FRW model}

In ADM formalism, the space-time ${\cal M }$ is assumed to be
foliated by a coordinate time $t$. Now, suppose that the metric
$g$ satisfies FRW dynamical equations which are assumed to include
a contribution from massless scalar field and we choose the
foliated 3-geometry,  $\Sigma(t)$ to be observer's past null
hypersurface and also the space-time contains a future-directed
time-like geodesic ${\Gamma}$ representing the world-line of an
observer.

Also suppose that the 4-volume time variable $T_\phi(t)$ defined
in (\ref{eq:cosmologicaltime}) instead of coordinate time $t$ has
been used to label the 3-surfaces and also the future-directed
time-like geodesic ${\Gamma}$ .

It is then possible to construct a covariantly defined geometric
quantity determined by field values on $\Sigma_{T_\phi}(t)$ \be
\Psi(\Sigma_{T_\phi})=\int_{\Sigma_{T_\phi}}
\psi(x)\sqrt{{}^{(3)}h} d^3 x,
 \ee
where  $\psi(x)$ is any scalar invariant on $\Sigma_{T_\phi}(t)$
expressible in terms of $h_{ij}$ , $R^i_{jkl}$, and their
covariant spatial derivatives.These quantities are called world
line ${\Gamma}$-observables  \cite{Hossein} for FRW model.

The so called ${\Gamma}$-observables then have vanishing poisson
brackets with any Hamiltonian $H$, equation
(\ref{eq:totalHamiltonian}), which generates time translations of
$\Sigma_{T_\phi}(t)$ along ${\Gamma}$. The observables
$\Psi(\Sigma_{T_\phi})$ do not have vanishing Poisson brackets
with the Hamiltonian constraint ${\cal H}_1$, since the
prespecified foliation is not invariant under local time evolution
\cite{Kuchar}.

If we define new quantities,$\Psi_{T_\phi}(\Sigma_{T_\phi})$ ; the
value $\Psi(\Sigma_{T_\phi})$ at a certain time $T_\phi$, then
these quantities have vanishing Poisson brackets with the
Hamiltonian constraint, $\{\Psi_{T_\phi}(\Sigma_{T_\phi}), {\cal
H}_1\}= 0$ , and can be called 'evolving constants of motion'.
These observables are the same as Rovelli's constants of motion in
a sense that they are genuine Dirac's observables. The evolution
of these observables is expressed in terms of the dynamical
variable $T_\phi(t)$, whose conjugate momenta, is a first class
constraint.Similarly, the dynamical time $T_\phi(t)$ in the new
labeling of 3-surfaces is not a Dirac observable although its
conjugate momenta is constant along the world line.

Alternatively, using (\ref{eq:cosmologicaltime1}) it is also
possible to construct a covariantly defined matter quantity
determined by the scalar factor values on $\Sigma_{T_a}(t)$ \be
\Psi(\Sigma_{T_a})=\int_{\Sigma_{T_a}} \psi(\phi)\sqrt{{}^{(3)}h}
d^3 x,
 \ee
where  $\psi(\phi)$ is any scalar invariant on $\Sigma_{T_a}(t)$
expressible in terms of $\phi$, and its covariant spatial
derivatives. These quantities are also called world line
${\Gamma}$-observables.

In summary we have seen that an explicit time variable has been
emerged in FRW model from gravity coupled to the massless scalar
field, interpreted as a cosmological time, and can be used by
observers as a clock to measure the passage of time. A set of
'evolving constants of motion' has been constructed by using the
dynamical time variable emerged from scalar field or scalar factor
which set the condition on the ${\Gamma}$-observables.

\section{ Acknowledgement}
I would like to thank Dr Hugh Luckock for his help in the
achievement of this work.


\newpage

\begin{thebibliography}{50}

\bibitem{Isham} C.J. Isham, 'Canonical Quantum Gravity and the
Problem of Time', Lecture presented at NATO Advanced Study
Institude "Recent problems in mathematical physics",
Salamanca(1992),gr-qc/9210011


\bibitem{Inverno}R. d'Inverno, 'Introducing Einstein's relativity', Clarendon
Press(1992). Oxford

\bibitem{Hawking}S.W.Hawking \& G.F.R. Ellis,' The large scale structure of
sapce-time', Cambridge University Press(1973)

\bibitem{Komar}A. Komar, Am.J. Phys.(1965)33,1024-1027

\bibitem{Dautcourt}G.Dautcourt,\& P.Babelsberg, 'Observer dependence of past light
cone initial data in cosmology',Astron.Nachr.,(1985)306, 1-6

\bibitem{hughhossein}H.Farajollahi,\& H.Luckock, 'Dirac
Observables and the Phase Space of General Relativity',General
Relativity and Gravitation,34(10):1685-1699,October(2002)

\bibitem{Penrose}R.Penrose, 'Null hypersurface initial data for classical fields of
arbitrary spin and for General Relativity', General Relativity and
Gravitation( 1980)12, 225-264

\bibitem{Sachs}R.K.Sachs, 'On the characteristic initial value problem in
gravitational theory',J.Math.Phys.(1962)3, 908-914

\bibitem{Ellis}G.F.R. Ellis, S.D. Nel, R. Maartens, W.R. Stoeger \& A.P. Whitman,
'Ideal observational cosmology', Phys.Rep.(1985)124,315-417

\bibitem{Bondi}H.Bondi, M. J. G. Van der Berg \& A.W.K. Metzner,'
Gravitational waves in General Relativity, VII Waves from
axi-symmetric isolated systems', Proc. Roy. Soc.
Lond.(1962)A269,21

\bibitem{palii} A.M. Khvedelidze, \& Yu. G. Palii, 'Generalized
Hamiltonian dynamics of Friedmann cosmology with scalar and spinor
matter source fields' Class. Quantum Grav. 18(2001)1767-1785

\bibitem{Dirac}Dirac P. A. M., 'Lectures on Quantum Mechanics',Belfer Graduate
School, New York,(1964)

\bibitem{Hossein}H. Farajollahi,'World-line observables and clocks in General
Relativity' To be published in 'International Journal of
Theoretical Physics, Group Theory and Nonlinear Optics'

\bibitem{Marolf}D. Marolf, 'Quantum observables and recollapsing dynamics', Class.
Quant. Grav.(1995)12, 1199

\bibitem{Rovelli}C.Rovelli,'What is observable in classical and quantum
gravity?',Class. Quantum Grav.(1991)8,297-316

\bibitem{Camelia}G. A. Camelia, 'On local observations in quantum gravity',Modern
Phys. Letters A(1996),11,No.17, 1411-1416

\bibitem{Kuchar}K. Kuchar, 'Canonical quantisation of gravity in Relativity,
Astrophysics and Cosmology', Reidel, Dordrecht,(1973) pp.237-288




\end{thebibliography}
\end{document}